\documentclass[a4paper,twoside,natbib]{article}
\newcommand{\affil}[1]{$^{\rm #1}$}
\usepackage[authoryear]{natbib}
\bibpunct{(}{)}{;}{a}{}{,}
\usepackage{graphicx}
\date{}

\title{\large\bf\flushleft Estimation of galactic model parameters in high latitudes with SDSS}
\author{\parbox{\textwidth}{\flushleft
\vspace{-0.5cm}
{\it S. Bilir \affil{A}, A. Cabrera-Lavers \affil{B,C}, S. Karaali\affil{D}, S. Ak\affil{A}, E. Yaz\affil{A}, and L\'opez-Corredoira\affil{B}\\
\vspace{0.4cm}
{\small \affil{A}\, Istanbul University Science Faculty, Department of Astronomy and Space Sciences, 34119, University-Istanbul, Turkey, Email: sbilir@istanbul.edu.tr}}\\
{\small \affil{B}\, Instituto de Astrof\'{\i}sica de Canarias, E-38205 La Laguna, Tenerife, Spain}\\
{\small \affil{C}\, GTC Project Office, E-38205 La Laguna, Tenerife, Spain}\\
{\small \affil{D}\, Beykent University, Faculty of Science and Letters, Department of Mathematics and Computing, Ayaza\u ga 34396, Istanbul, Turkey}\\
}}
\begin{document}
\twocolumn[
\begin{changemargin}{.8cm}{.5cm}
\begin{minipage}{.9\textwidth}
\vspace{-1cm}
\maketitle
\small{\bf Abstract:}
We estimated the Galactic model parameters for a set of 36 high-latitude fields included in the currently available Data Release 5 (DR 5) of the
Sloan Digital Sky Survey ({\em SDSS\/}), to explore their possible variation with the Galactic longitude. The thick disc scaleheight moves from $\sim$550 pc at $120^{\circ}<l<150^{\circ}$ to $\sim$720 pc at $250^{\circ}<l<290^{\circ}$, while the thin disc scaleheight is as large as $\sim$195 pc in the anticenter direction and $\sim$15\% lower at $|l|<30^\circ$. Finally, the axis ratio ($c/a$) of the halo changes from a mean value of $\sim$0.55 in the two first quadrants of the Galaxy to $\sim$0.70 at $190^{\circ}<l<300^{\circ}$. For the halo, the reason for the dependence of the model parameters on the Galactic longitude arises from the well known asymmetric structure of this component. However, the variation of the model parameters of the thin and thick discs with Galactic longitude originates from the gravitational effect of the Galactic long bar. Moreover, the excess of stars in quadrant I (quadrant III) over quadrant IV (quadrant II) is in agreement with this scenario.\\
\medskip{\bf Keywords:} Galaxy: disc, Galaxy: structure, Galaxy: fundamental parameters
\medskip
\medskip
\end{minipage}
\end{changemargin}
]
\small

\section{Introduction}

The traditional star-count analysis of the Galactic structure have provided a picture of the basic structural and stellar populations of the Galaxy. Examples and reviews of these analyses can be found in \cite{Bahcall86}, \cite{GWK89}, \cite{Majewski93}, \cite{Robin00} and recently \cite{Chen01} and \cite{Siegel02}. The largest of the observational studies prior to the {\em SDSS\/} \citep{York00} are based on photographic surveys. The Basle Halo Program has presented the largest systematic photometric survey of the Galaxy \citep{Becker65, F89a,F89b,F89c,F89d}. The Basle Halo Program photometry is currently being re-calibrated and re-analysed, using an improved calibration of the {\em RGU\/} photometric system \citep{BF90, Buser98, Buser99}. More recent and future studies are being based on charge-coupled device (CCD) survey data. 

Our knowledge of the structure of the Galaxy, as inferred from star count data with colour information, entered now to the next level of precision with the advent of new surveys such as {\em SDSS\/}, {\em 2MASS\/}, {\em CADIS}, {\em BATC}, {\em DENIS\/}, {\em UKIDSS/VISTA}, {\em CFH/Megacam\/}, and {\em Suprime\/}. Researchers have used different methods to determine the Galactic model parameters. In Table 1 of \cite{KBH04} we can find an exhaustive list of the different values obtained for the structural parameters of the discs and halo of the Milky Way. One can see directly that there is a refinement in the numerical values of the model parameters. The local space density and the scaleheight of the thick disc can be given as an example. The evaluation of the thick disc have steadily moved towards shorter scaleheights, from 1.45 to 0.65 kpc \citep{GR83, Chen01}, and higher local densities (2--10\%). In many studies the range of values for the parameters is large. For example, \cite{Chen01} and \cite{Siegel02} give 6.5--13 and 6--10\%, respectively, for the local space density for the thick disc. However, one expects the most accurate numerical values for these recent works. That is, either the range for the parameters should be small or a single value with a small error should be given for each of them. It seems that workers have not been able to choose the most appropriate procedures in this topic.

Large range or different numerical values for a specific Galactic model parameter as estimated by different researchers may be  due to several reasons: 1) The Galactic model parameters are Galactic latitude / longitude dependent. The two works of \cite{Buser98, Buser99} confirm this suggestion. Although these authors give a mean value for each parameter, there are differences between the values of a given parameter for different fields. Also, it has been recently shown that the Galactic model parameters are Galactic longitude dependent \citep{Bilir06a, Bilir06b, Cabrera07, Ak07a}. 2) The Galactic model parameters are absolute magnitude dependent \citep{KBH04, Bilir06c}. Hence, any procedure which excludes this argument give Galactic model parameters spread in a large range. 3) Galactic model parameters change with limiting distance of completeness. That is, a specific model parameter is not the same for each set of Galactic model parameters estimated for different volumes \citep{Karaali07}.  

The difference between the Galactic model parameters estimated for fields with different Galactic latitudes and longitudes can be explained by the influence of the disc flaring and warping. The disc of our Galaxy is far from being radially smooth and uniform. On the contrary, its overall shape presents strong asymmetries.While the warp bends the Galactic plane upwards in the first and second Galactic longitude quadrants (0$^\circ\le l \le180^\circ$) and  downwards in the third and fourth quadrants (180$^\circ\le l \le360^\circ$), the flare changes the scaleheight as a function of radial distance. 

This warp is present in all Galactic components: dust \citep{DS01, Mar06}, gas \citep{Burton88, DS01, NS03, Levine06, VB06}, and stars \citep{L02, Momany06}. All these components have the same node position, and their distributions are asymmetric. However, the amplitude of the dust warp seems to be less pronounced than the stellar and gaseous warps, that share the same approximate amplitude \citep{L02, Momany06}. 
 
The stellar and gaseous flarings for the Milky Way are also compatible \citep{Momany06}, showing that $h_z$ increases with the galactocentric radius for $R>5$ kpc \citep{Kent91, DS01, NJ02, L02, Momany06}. The behaviour of this flare in the central discs of spiral galaxies has not been studied so well due to inherent difficulties in separating the several contributions to the observed counts or flux. \cite{L04}, for example, find that there is a deficit of stars compared to the predictions of a pure exponential law in the inner 4 kpc of the Milky Way, which could be explained as being a flare which displaces the stars to higher heights above the plane as we move to the Galactic centre.

In this scenario, where on the one hand the mean disc ($z=0$) can be displaced as much as 2 kpc between the location of the maximum and the minimum amplitudes of the warp \citep{DS01, L02, Momany06}, and on the other the scaleheight of the stars can show differences up to 50\% of the value for $h_z(R_{\odot})$ in the range $5<R<10$ kpc \citep{Al00, L02, Momany06} to fit a global Galactic disc model that accounts for all these inhomogeneities is, at the very least, tricky. Because of this the results in the Galactic model parameters might depend on the sample of Galactic coordinates used, as the combined effect of the warp and flare will be different at different directions in the Galaxy, hence at different lines of sight.

There is an additional reason for explaining the differences between the numerical values of a given Galactic model parameter estimated in different direction of the Galaxy, mainly at larger galactocentric distances. This is the observed overdensity regions with respect to an axisymmetric halo, where two competing scenarios have been proposed for its explanation: the first one is concerned with the triaxiality of the halo \citep{Newberg06, Xu06, J08} whereas the second one is related to the remnants of some historical merger events \citep{Wyse05}.

In this paper, we derive the structural parameters of two discs and halo of the Galaxy from very recent {\em SDSS\/} data to observe possible changes in the parameters with the Galactic longitude. We used about 1.27 million stars in 36 high-latitude fields which cover the whole longitude interval (0$^\circ\le l \le360^\circ$), and we evaluated their absolute magnitudes by means of the recent procedures which provides accurate distance determination. In Sect. 2 we describe the {\em SDSS\/} data, as well as the density laws, absolute magnitudes, distances, and density functions employed in the analysis. Estimation of the Galactic model parameters and their dependence with the Galactic longitude is given in Sect. 3. Finally, our main results are discussed and summarised in Sect. 4 and 5, respectively. 

\section{SDSS}
The {\em SDSS\/} is a large, international collaboration project set up to survey 10 000 square--degrees of sky in five optical passbands and to obtain spectra of one million galaxies, 100 000 quasars, and tens of thousands of Galactic stars. The data are being taken with a dedicated 2.5-m telescope located at Apache Point Observatory (APO), New Mexico. The telescope has two instruments: a CCD camera with 30 2048$\times$2048 CCDs in 
the focal plane and two 320 fiber double spectrographs. The imaging data are tied to a network of brighter astrometric standards (which would be saturated in the main imaging data) through a set of 22 smaller CCDs in the focal plane of the imaging camera. An 0.5-m telescope at APO will be used to tie the imaging data to brighter photometric standards. 

The {\em SDSS\/} obtains images almost simultaneously in five broad bands ($u$, $g$, $r$, $i$ and $z$)\footnote {Magnitudes in this paper are quoted in the $ugriz$ system to differentiate them from the former one, $u^{'}g^{'}r^{'}i^{'}z^{'}$} centred at 3551, 4686, 6166, 7480 and 8932 \AA,  respectively \citep{Fukugita96}. The imaging data are automatically processed through a series of software pipelines which find and measure objects and provide photometric and astrometric calibrations to produce a catalogue of objects with calibrated magnitudes, positions and structure information. The photometric pipeline \citep{Lupton01} detects the objects, matches the data from the five filters, and measures instrumental fluxes, positions, and shape parameters (which allows the classification of objects as ``point source'', -compatible with the point spread function-, or ``extended''). The photometric calibration is accurate to roughly 2\% rms in the $g$, $r$ and $i$ bands, and 3\% in $u$ and $z$, as determined by the constancy of stellar population colours \citep{Ivezic04, Blanton05}, while the astrometric calibration precision is better than 0.1 arcsec rms per coordinate \citep{Pier03}. The Data Release 5 (DR5) imaging catalogue covers 8000 deg$^{2}$ \citep{A07} with a detection repeatability complete at a 95\% level for point sources brighter than the limiting apparent magnitudes of 22.0, 22.2, 22.2, 21.3 and 20.5 mag for $u$, $g$, $r$, $i$ and $z$, respectively. The data are saturated at about 14 mag in $g$, $r$ and $i$ and about 12 mag in $u$ and $z$.  

\subsection{Observational data and reduction}

\begin{figure}
\begin{center}
\includegraphics[angle=0, width=75mm, height=39.9mm]{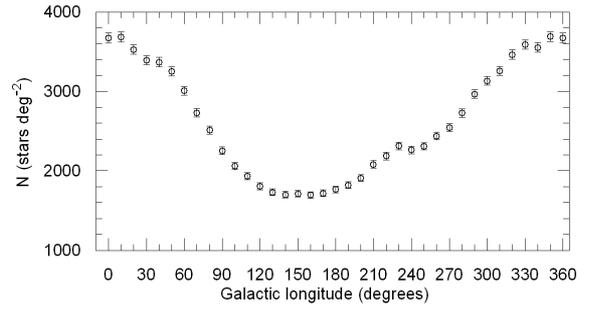}
\caption[] {Star counts at $b$=+62$^{\circ}$.5 for the 36 fields available in the DR5.} 
\label{SDSS_starcount}
\end{center}
\end{figure}

The data used in this work were taken from {\em SDSS\/} (DR5) WEB server\footnote{http://www.sdss.org/dr5/access/index.html} for 36 high-latitude fields ($60^\circ\leq b \leq65^\circ$) covering different Galactic longitude intervals ($0^\circ\leq l \leq360^\circ$). {\em SDSS\/} magnitudes $u$, $g$, $r$, $i$, and $z$ were used in a total number of 2 164 680 stars in 36 fields equal in size (totally 831 deg$^{2}$). Although the fields are equal in size, (23.08 deg$^2$) their surface densities (number of stars per square-degree) are not the same, following a specific trend with Galactic longitude (Fig. \ref{SDSS_starcount}). This is the first clue for the dependence of the Galactic model parameters with the Galactic longitudes. Owing to the {\em SDSS} observing strategy, stars brighter than $g_{o}=14^{m}$ are saturated, and star counts are not be complete for magnitudes fainter than $g_{o}=22^{m}.2$. Hence, our work is restricted to the magnitude range $15<g_{0}\leq22$ for the evaluation of the Galactic model parameters. 

The $E(B-V)$ colour excess was evaluated individually for each subsample source by using the maps of \cite{Schlegal98} through {\em SDSS\/} query server and this was reduced to total absorption $A_{V}$ via Eq. 1
\begin{eqnarray}
A_{V}=3.1E(B-V).
\end{eqnarray}
In order to determine total absorptions, $A_{m}$, for the {\em SDSS\/} bands, $A_{m}/A_{V}$ data given by \cite{Fan99}, i.e. 1.593, 1.199, 0.858, 0.639 and 0.459 for $m$ = $u$, $g$, $r$, $i$ and $z$ were used respectively. Thus, the de-reddened magnitudes, with subscript 0, are  
 
\begin{eqnarray}
u_{0}=u-A_{u},\\
g_{0}=g-A_{g},\\
r_{0}=r-A_{r},\\
i_{0}=i-A_{i},\\
z_{0}=z-A_{z}.
\end{eqnarray}
The total absorptions $A_{m}$ are avaliable in the {\em SDSS\/} query server.

All the colours and magnitudes mentioned hereafter will be de-reddened ones. Given that the location of the vast majority of our targets are at distances larger than 0.4 kpc, it seems appropriate to apply the full extinction from the maps. Actually, when we combine the distance $r=0.4$ kpc (distance 0.35 kpc from the Galactic plane) with the scaleheight of the dust, \citep[$H=125$ pc]{Mar06}, we find that the total extinction is reduced to 6\% of the value at the Galactic plane, at the nearest distance of stars in our work.

\begin{figure}
\begin{center}
\includegraphics[scale=0.31, angle=0]{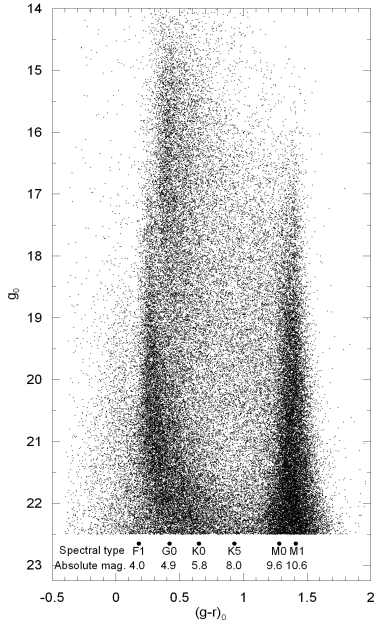}
\caption[] {Colour magnitude diagram for the star sample. Spectral types and absolute magnitudes are indicated in the horizontal axis. Then thin disc stars are dominant at the red peak whereas the thick disc ($14<g_{0}\leq18$) and the halo ($g_{0}>18$) stars are concentrated at the blue peak.} 
\label{cmd}
\end{center}
\end{figure}

According to \cite{Chen01}, the distribution of stars in the $g_{o}/(g-r)_{o}$ colour-magnitude diagram (CMD) can be classified as follows: 
Blue stars in the range magnitude $15<g_{0}<18$ are dominated by thick-disc stars with a turn-off at $(g-r)_{0}\approx 0.33$ mag, while Galactic halo stars
become significant for $g_{0}>18$ mag, with a turn-off at $(g-r)_{0} \approx 0.2$ mag. Red stars, $(g-r)_{0} \geq 1.3$ mag, are dominated by thin disc stars at all apparent magnitudes. The CMD, $g_{0}/(g-r)_{0}$, in Fig. \ref{cmd} shows the mentioned populations and, spectral types and absolutes magnitude of stars of these populations.  

\begin{figure}
\begin{center}
\includegraphics[angle=0, width=70mm, height=79mm]{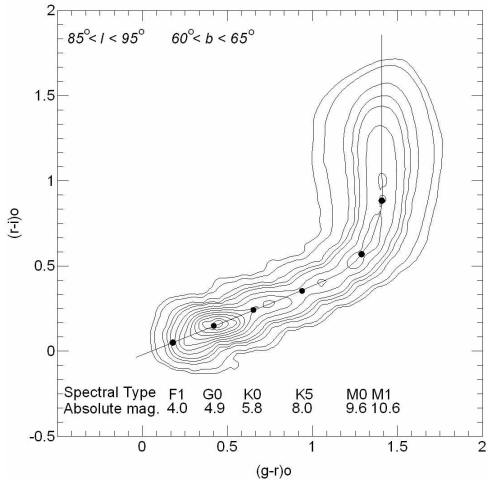}
\caption[] {$(g-r)_{0}/(r-i)_{0}$ two colour diagram for the field centred at $l=90^\circ$. Isodensity contours show the position of
stars at distance $d<0.3$ mag. from the stellar locus, adopted from \cite{J08}. Black circles represent the dwarf stars with spectral
types and absolute magnitudes stated in two lines below the diagram.} 
\label{gr-ri}
\end{center}
\end{figure}

However, the CMDs and the two-colour diagrams for all objects (not presented here) indicate that the stellar distributions are contaminated by extragalactic objects as claimed by \cite{Chen01}. The star/extragalactic object separation is based on the ``stellarity parameter'' as returned from the SE{\tiny XTRACTOR} routines \citep{Bertin96}. This parameter has a value between 0 (high extended) and 1 (point source). The separation works very well to classify a point source with a value greater than 0.8. Needless to say, this separation depends strongly on seeing and sky brightness. We also applied the ``locus-projection'' method of \cite{J08} in order to remove hot white dwarfs, low-redshift quasars, and white/red dwarf unresolved binaries from our sample. Briefly, this procedure consists of rejecting objects at distances larger than 0.3 mag from the stellar locus (Fig. \ref{gr-ri}).

\subsection{Density laws}

In this work we adopted the density laws of the Basle group \citep{Buser98, Buser99}. Disc structures are usually parameterized in 
cylindrical coordinates by radial and vertical exponentials:
\begin{equation}
D_{i}(R,z)=n_{i}~\exp(-|z|/h_{z,i})~\exp(-(R-R_{0})/h_{i}),\\
\label{ec1}
\end{equation}
where $z=z_{\odot}+r\sin(b)$, $r$ is the distance to the object from the Sun, $b$ the Galactic latitude, $z_{\odot}$  the vertical
distance of the Sun from the Galactic plane \citep[24 pc]{J08}, $R$ the projection of the galactocentric distance on the Galactic plane, $R_{0}$ the solar distance from the Galactic centre \citep[8 kpc]{R93}, $h_{z,i}$ and $h_{i}$ are the scaleheight and scalelength, respectively, and $n_{i}$ is the normalized density at the solar radius. The suffix $i$ takes the values 1 and 2 as long as the thin and thick discs are considered. As this study focuses on the dependence of the scaleheight and solar normalization on the Galactic longitude, we fixed their scalelengths to 2.4 and 3.5 kpc for thin and thick discs, respectively, according to \cite{J08}. 

The density law for the spheroid component is parameterized in different forms. The most common is the \cite{de48} spheroid used to describe the 
surface brightness profile of elliptical galaxies. This law has been deprojected into three dimensions by \cite{Young76} as 
\begin{eqnarray}
D_{s}(R)=n_{s}~\exp[-7.669(R/R_{e})^{1/4}]/(R/R_{e})^{7/8},
\end{eqnarray}
where $R$ is the (uncorrected) Galactocentric distance in spherical coordinates, $R_{e}$ is the effective radius and $n_{s}$ is the normalized 
local density. $R$ has to be corrected for the axial ratio $\kappa = c/a$, 
\begin{eqnarray}
R = [x^{2}+(z/\kappa)^2]^{1/2},
\end{eqnarray}
\begin{eqnarray}
z = z_{\odot}+ r \sin b,
\end{eqnarray}
\begin{eqnarray}
x = [R_{o}^{2}+(z/\tan b)^2-2R_{o}(z/\tan b)\cos l]^{1/2},
\end{eqnarray}
with $r$ the distance along the line of sight and, $b$ and $l$ the Galactic latitude and longitude respectively, for the field under investigation. The form used by Basle group \citep{F89a,F89b,F89c,F89d} is independent of effective radius but is dependent on the distance from the Sun to the 
Galactic centre,  
\begin{eqnarray}
D_{s}(R)=n_{s}\exp[10.093(1-(R/R_{o})^{1/4})]/(R/R_{o})^{7/8}.
\end{eqnarray}

\subsection{Absolute magnitudes, distances, and density functions}
Absolute magnitudes were determined by two different procedures. For absolute magnitudes $4<M_{g}\leq8$ we used the procedure of \cite{KBT05}, whereas for $8<M_{g}\leq10$ we adopted the procedure of \cite{Bilir05}. The cited absolute magnitude intervals correspond to the spectral type intervals F0--K5 and K5--M0 respectively. In the procedure of KBT, the absolute magnitude offset from the Hyades main sequence, $\Delta M_{g}^{H}$, is given as a function of both $(g-r)_{0}$ colour and $\delta_{0.43}$ UV-excess, as follows:

\begin{eqnarray}
\Delta M_{g}^{H}=c_{3}\delta^{3}_{0.43}+c_{2}\delta^{2}_{0.43}+c_{1}\delta_{0.43}+c_{0},
\label{eq7}
\end{eqnarray}
where $\delta_{0.43}$ is the UV-excess of a star relative to a Hyades star of the colour-index $(g-r)_{0}=0.43$ which corresponds to $\delta_{0.6}$ and which is determined by the colour equations between {\em UBV} and {\em SDSS} photometry (KBT). The coefficients $c_{i}$ (i=0, 1, 2, 3) are functions of $(g-r)_{0}$ colour and they are adopted from the work of KBT, and where $\Delta M^{H}_{g}$ is defined as the difference in absolute magnitudes of a program star and a Hyades star of the same $(g-r)_{0}$ colour:

\begin{eqnarray}
\Delta M^{H}_{g} = M^{*}_{g}-M^{H}_{g},
\label{eq8}
\end{eqnarray}
The absolute magnitude for a Hyades star can be evaluated from the Hyades sequence, normalized by KBT (Eq. 15 of KBT). This procedure is the one used in the work of \cite{Ak07b}, and has two main advantages: 1) there is no need to separate the stars into different populations, and 2) the absolute magnitude of a star is determined from its UV-excess individually which provides more accurate absolute magnitudes compared with the procedure 'in-situ', where a specific CMD is used for all stars of the same population. When one uses the last two equations (Eqs. \ref{eq7} and \ref{eq8}) and Eq. 15 of KBT simultaneously, it gets the absolute magnitude $M_{g}^{*}$ of a star. 

The procedure of KBT was defined only for the colours $0.09<(g-r)_{0}\leq0.93$, which corresponds to absolute magnitudes $4<M_{g}\leq8$. Hence, for the absolute magnitudes $8<M_{g}\leq10$ we used the equation of BKT, which provides absolute magnitudes for late-type dwarfs:
 
\begin{eqnarray}
M_{g} = 5.791 (g-r)_{0} + 1.242 (r-i)_{0} + 1.412.
\end{eqnarray}

Stars with faint absolute magnitudes are very useful, since they provide estimates for the space densities at short distances relative to the Sun which combine the space densities for absolutely bright stars at large distances, and the local densities of {\em Hipparcos} \citep{Jahreiss97}. Thus, we have a sample of stars with absolute magnitudes $4<M_{g}\leq10$ which enables us to evaluate space density functions in the heliocentric distances interval $0.4<r\leq25$ kpc, corresponding to a range of distances of $0.4<z\leq21$ kpc from the Galactic plane. This interval is large enough to estimate a set of Galactic model parameters and test their change with the Galactic longitude. The absolute magnitudes in question and the corresponding spectral types (from early F type to early M type) for the locus points in the $(g-r)_{0}/(r-i)_{0}$ two colour diagram is shown in Fig. \ref{gr-ri} for the field centred at $l=90^\circ$ as an example. The local space density in the interval $4<M_{g}\leq10$ is flat and it attributes a mean value of logarithmic space density $D^{*}=7.49$.

In a conical magnitude-limited volume, the distance to which intrinsically bright stars are visible is larger than the distance to which intrinsically faint stars are visible. The effect of this is that brighter stars are statistically overrepresented and the derived absolute magnitudes are too faint. This effect, known as Malmquist bias \citep{M20}, was formalized into the general formula:
\begin{equation}
M_{g}=M_{0}-\sigma^{2}{d\log A(g) \over dg},
\end{equation}
where $M_{g}$ is the assumed absolute magnitude, $M_{0}$ is the absolute magnitude calculated for any star using KBT calibration, $\sigma$ is the dispersion of the KBT or BKT calibration, and $A(g)$ is the differential counts evaluated at the apparent magnitude $g_{0}$ of any star. The dispersion in absolute magnitude calibration of KBT and BKT is around 0$^{m}$.25, corresponding an error about 10\% in photometric distance.  We divided stars into the absolute magnitude intervals (4,5], (5,6], (6,7], (7,8], (8,9] and (9,10], and we applied the Malmquist bias to stars in each interval separately. This approach provides (relative) uniform space densities which is the essential Malmquist bias. Thus, the corrections applied to the absolute magnitudes are 0.005, 0.003, 0.007, 0.008, 0.012 and 0.012 for the absolute magnitude intervals cited above. The correction of the Malmquist bias was applied to the {\em SDSS} photometric data used in this work.

The combination of the absolute magnitude $M_{g}$ and the apparent magnitude $g_{0}$ of a star gives its distance $r$ relative to the Sun, i.e.,

\begin{equation}
[g-M_{g}]_{0}=5\log r-5.
\end{equation}
\cite{GWJ95} quote an error of $\sim$0.2 dex in the derivation of $[M/H]$ from the {\em UBV} photometry for F/G stars which leads to a random uncertainty of $20\%$ in the distance estimation. One expects larger distance errors for later spectral type stars. However, $(u-g)$ and $(g-r)$ colours are more accurate than the $(U-B)$ and $(B-V)$ colours which compensates this excess error for K stars.  

\begin{figure}
\begin{center}
\includegraphics[angle=0, width=75mm, height=52.5mm]{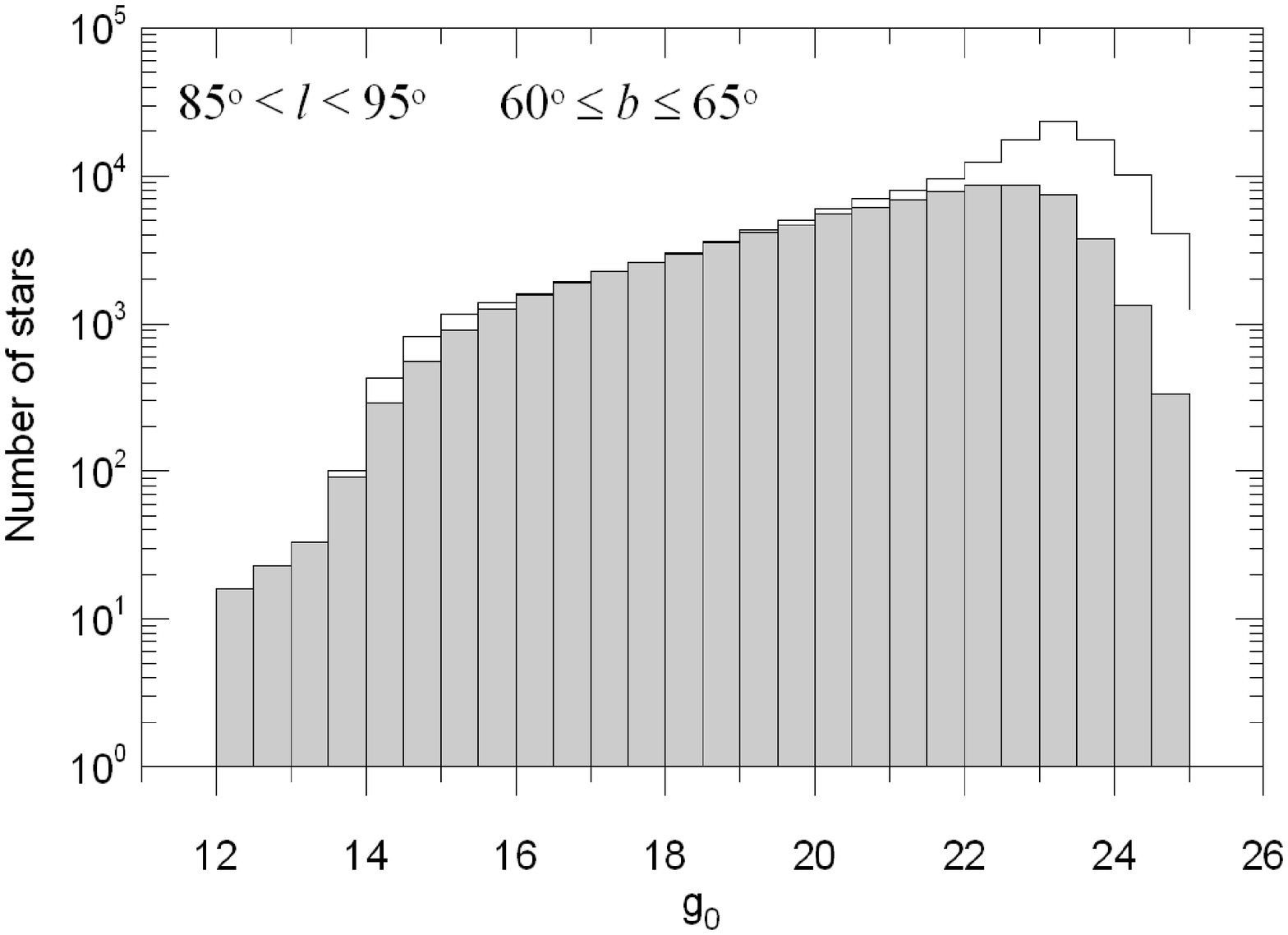}
\caption[] {Apparent magnitude histogram for point sources (white area) and for final stars sample (shaded area) for the field 
centred at $l=90^\circ$.}
\label{star-sample}
\end{center}
\end{figure}

Logarithmic space densities $D^{*}=\log D+10$ have been evaluated for the combination of three population components (thin and thick discs and
halo), for each field where $D=N/ \Delta V_{1,2}$; $\Delta V_{1,2}=(\pi/180)^{2}(A/3)(r_{2}^{3}-r_{1}^{3})$; $A$ denotes the size of the
field (23.08 deg$^{2}$); $r_{1}$ and $r_{2}$ are the lower and upper limiting distances of the volume $\Delta V_{1,2}$; $N$ is the number
of stars per unit absolute magnitude; $r^{*}=[(r^{3}_{1}+r^{3}_{2})/2]^{1/3}$ is the centroid distance of the volume $\Delta V_{1,2}$; and
$z^{*}=r^{*}\sin(b)$, $b$ being the Galactic latitude of the field centre. The limiting distances of completeness, $z_{l},$ can be  calculated from the following equations:

\begin{eqnarray}
[g_{l}-M_{g}]_{0} =  5 \log r_{l} - 5,\\
z_{l}=r_{l} \sin (b),					
\end{eqnarray}
where $g_{l}$ is the limiting apparent magnitude (15 and 22, for the bright and faint stars, respectively, Fig. \ref{star-sample}), $r_{l}$ is the limiting distance of completeness relative to the Sun, and $M_{g}$ corresponds the absolute magnitude defining the interval $(M_{1},M_{2}]$ where $(M_{1},M_{2})$ is (4,5], (5,6], (6,7], (7,8], (8,9] and (9,10]. For the limiting distance of completeness at short and large distances $M_{g}$ assumes the bright and faint absolute magnitudes for each absolute magnitude interval. That is, the limiting distance of completeness is defined for each absolute magnitude to a limited range of spectral types.

\begin{figure*}
\begin{center}
\includegraphics[angle=0, width=160mm, height=84mm]{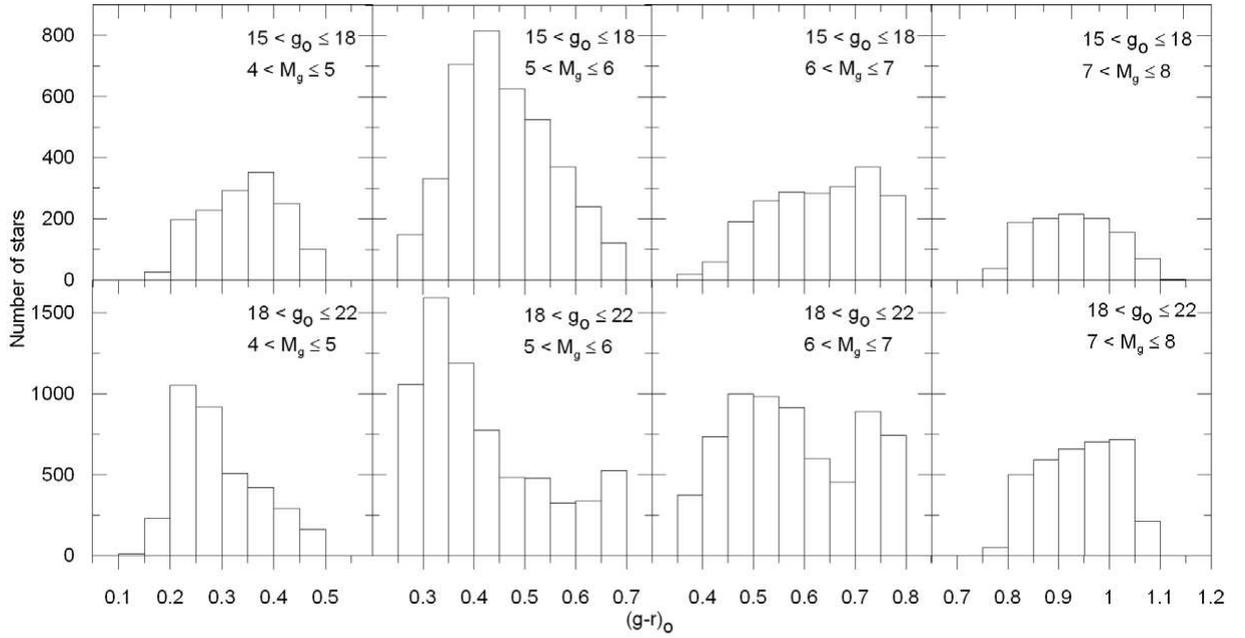}
\caption[] {The $(g-r)_{0}$ colour distribution as a function of apparent and absolute magnitudes, for the field centred at $l=90^\circ$.} 
\label{his:col}
\end{center}
\end {figure*}

We present the distribution of $(g-r)_{0}$ colours and $[M/H]$ metallicities for stars in the field centred at $l=90^\circ$, as an example, to show the variation in these parameters as a function of apparent $g_{0}$ and absolute $M_{g}$ magnitudes. Fig. \ref{his:col} shows how for the same apparent magnitude interval the peaks of the histograms move to redder colours when one goes to fainter absolute magnitudes. This result is in agreement with the fact that stars closer to the Sun are late-type stars, i.e. thin disc stars. Another result is that the peaks of the histograms for stars with fainter apparent magnitudes ($18<g_{0}\leq22$), but with the same absolute magnitudes as the brighter ones, occupy bluer colours. This confirms the suggestion of \cite{Chen01}, who demonstrated in their Fig. 6 that apparently fainter stars ($g_{0}>18$) are dominated by blue stars, i.e. halo stars. 

\begin{figure*}
\begin{center}
\includegraphics[angle=0, width=160mm, height=84mm]{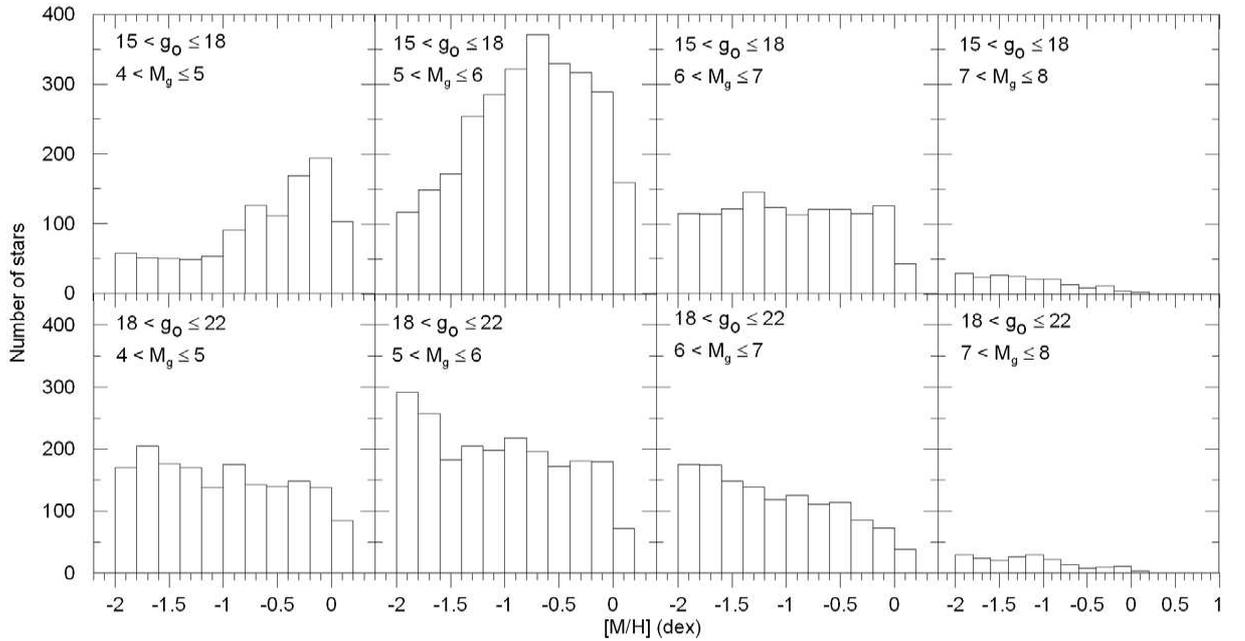}
\caption[] {The metallicity distribution as a function of apparent and absolute magnitudes, for the field centred at $l=90^\circ$.} 
\label{his:met}
\end{center}
\end {figure*}

Additionally, Fig. \ref{his:met} shows three peaks (at metal-rich, intermediate and metal-poor parts) in the metallicity distribution. However, all of them are not conspicuous in the same panel, as they correspond to different absolute and apparent magnitude intervals. If we compare the upper and lower panels for stars with the same absolute magnitudes but different apparent magnitudes, we notice that the peaks shift to lower metallicities when one goes from relatively bright to faint apparent magnitude intervals. As different peaks mean different components of our Galaxy, i.e. thin and thick discs and halo, the result stated above confirms our previous finding \citep{Karaali07} that different populations are dominant at different absolute magnitudes. Most of the stars with fainter absolute magnitudes, $6<M_{g}\leq7$ for example, are metal rich stars (disc stars), whereas the absolutely brightest stars, i.e. $4<M_{g}\leq5$, have low metallicities, $[M/H]<-1$ dex, and hence belong to the halo component of our Galaxy. As \citet{Carney79} quoted, the UV-excess at the red end is small which limits the accuracy of the metallicity estimation (see Fig. \ref{two-colour} in the Appendix). However, the general trend of metallicity distribution in Fig. \ref{his:met} does not give the indication of such an effect.

\begin{table*}
\center
\caption{Galactic model parameters for 36 {\em SDSS} high-latitude ($60^\circ\leq b \leq65^\circ$) fields, resulting from the fits of the analytical density profiles. The columns indicate: Galactic longitude ($l$), scaleheight of thin ($h_{z,1}$) and thick discs ($h_{z,2}$), local space density of the thick disc $(n_{2}/n_{1})$ and the halo $(n_{3}/n_{1})$ relative to the thin disc, axial ratio of the halo ($c/a$) and star number density of the field ($N$), reduced chi-–square minimum ($\widetilde{\chi}^{2}_{min}$), and the corresponding probability.}
\label{tableSDSS}
\begin{tabular}{ccccccccc}
\hline
    & Thin disc & \multicolumn{2}{c}{Thick disc}&\multicolumn{2}{c}{Halo} & & & \\
\hline
$<l>$  & $h_{z,1}$  & $h_{z,2}$  & $n_{2}/n_{1}$   & $n_{3}/n_{1}$  &  $c/a$  & $N$ & $\widetilde{\chi}^{2}_{min}$ & Probability\\
($^\circ$) &  (pc) &  (pc) & $(\%)$   & $(\%)$  &    & $(stars/deg^{2})$ &  & \\
\hline
  0 & 177$\pm$9  & 634$\pm$46 & 10.96$\pm$1.23 &  0.14$\pm$0.01 & 0.60$\pm$0.02 & 3672$\pm$61 & 0.517 & 0.972\\
 10 & 178$\pm$6  & 613$\pm$28 & 12.50$\pm$0.90 &  0.16$\pm$0.01 & 0.58$\pm$0.01 & 3686$\pm$61 & 0.488 & 0.981\\
 20 & 167$\pm$8  & 586$\pm$27 & 15.14$\pm$1.15 &  0.17$\pm$0.01 & 0.57$\pm$0.01 & 3527$\pm$59 & 0.438 & 0.991\\
 30 & 187$\pm$9  & 620$\pm$44 & 11.67$\pm$1.33 &  0.17$\pm$0.01 & 0.56$\pm$0.01 & 3391$\pm$58 & 0.844 & 0.677\\
 40 & 180$\pm$6  & 595$\pm$28 & 13.46$\pm$1.46 &  0.19$\pm$0.02 & 0.54$\pm$0.02 & 3370$\pm$58 & 0.475 & 0.984\\
 50 & 172$\pm$6  & 609$\pm$26 & 13.30$\pm$0.89 &  0.16$\pm$0.01 & 0.58$\pm$0.02 & 3251$\pm$57 & 0.537 & 0.965\\
 60 & 195$\pm$6  & 641$\pm$35 & 10.33$\pm$0.86 &  0.13$\pm$0.01 & 0.62$\pm$0.01 & 3008$\pm$55 & 0.795 & 0.742\\
 70 & 181$\pm$7  & 616$\pm$38 & 11.07$\pm$1.01 &  0.12$\pm$0.01 & 0.62$\pm$0.01 & 2733$\pm$52 & 0.335 & 0.999\\
 80 & 178$\pm$7  & 598$\pm$35 & 10.86$\pm$1.03 &  0.14$\pm$0.01 & 0.58$\pm$0.01 & 2514$\pm$50 & 0.399 & 0.995\\
 90 & 192$\pm$7  & 647$\pm$52 &  7.89$\pm$1.02 &  0.12$\pm$0.01 & 0.60$\pm$0.02 & 2252$\pm$47 & 0.541 & 0.963\\
100 & 189$\pm$5  & 638$\pm$35 &  7.94$\pm$0.69 &  0.12$\pm$0.01 & 0.58$\pm$0.01 & 2063$\pm$45 & 0.310 & 0.999\\
110 & 186$\pm$4  & 648$\pm$35 &  7.11$\pm$0.62 &  0.12$\pm$0.01 & 0.59$\pm$0.01 & 1932$\pm$44 & 0.479 & 0.983\\
120 & 196$\pm$7  & 667$\pm$68 &  5.45$\pm$0.90 &  0.15$\pm$0.01 & 0.54$\pm$0.01 & 1807$\pm$43 & 0.855 & 0.662\\
130 & 184$\pm$6  & 581$\pm$36 &  8.75$\pm$0.96 &  0.16$\pm$0.01 & 0.54$\pm$0.01 & 1732$\pm$42 & 0.468 & 0.986\\
140 & 172$\pm$4  & 550$\pm$23 & 10.26$\pm$0.70 &  0.16$\pm$0.01 & 0.54$\pm$0.01 & 1698$\pm$41 & 0.413 & 0.994\\
150 & 183$\pm$4  & 568$\pm$26 &  8.71$\pm$0.69 &  0.16$\pm$0.01 & 0.54$\pm$0.01 & 1712$\pm$41 & 0.450 & 0.989\\
160 & 195$\pm$8  & 603$\pm$64 &  7.23$\pm$1.22 &  0.16$\pm$0.03 & 0.53$\pm$0.04 & 1697$\pm$41 & 0.543 & 0.962\\
170 & 195$\pm$7  & 638$\pm$67 &  6.35$\pm$1.05 &  0.17$\pm$0.01 & 0.54$\pm$0.01 & 1719$\pm$41 & 0.816 & 0.715\\
180 & 195$\pm$4  & 658$\pm$42 &  6.22$\pm$0.62 &  0.14$\pm$0.01 & 0.61$\pm$0.03 & 1766$\pm$42 & 0.681 & 0.869\\
190 & 198$\pm$4  & 646$\pm$37 &  6.46$\pm$0.59 &  0.12$\pm$0.01 & 0.65$\pm$0.01 & 1822$\pm$43 & 0.630 & 0.912\\
200 & 199$\pm$7  & 657$\pm$61 &  6.35$\pm$0.91 &  0.13$\pm$0.01 & 0.67$\pm$0.03 & 1911$\pm$44 & 0.541 & 0.963\\
210 & 179$\pm$5  & 598$\pm$31 &  9.46$\pm$0.77 &  0.14$\pm$0.01 & 0.71$\pm$0.02 & 2083$\pm$46 & 0.396 & 0.996\\
220 & 185$\pm$7  & 594$\pm$46 &  9.16$\pm$1.12 &  0.13$\pm$0.01 & 0.76$\pm$0.02 & 2186$\pm$47 & 0.656 & 0.892\\
230 & 198$\pm$4  & 654$\pm$36 &  7.08$\pm$0.59 &  0.13$\pm$0.01 & 0.75$\pm$0.01 & 2314$\pm$48 & 0.372 & 0.997\\
240 & 174$\pm$4  & 603$\pm$26 &  9.18$\pm$0.57 &  0.13$\pm$0.01 & 0.71$\pm$0.04 & 2264$\pm$48 & 0.495 & 0.979\\
250 & 189$\pm$4  & 621$\pm$27 &  8.09$\pm$0.56 &  0.16$\pm$0.01 & 0.64$\pm$0.01 & 2310$\pm$48 & 0.331 & 0.999\\
260 & 188$\pm$6  & 658$\pm$44 &  7.53$\pm$0.79 &  0.12$\pm$0.01 & 0.72$\pm$0.02 & 2439$\pm$49 & 0.538 & 0.964\\
270 & 186$\pm$9  & 684$\pm$80 &  6.82$\pm$1.23 &  0.11$\pm$0.01 & 0.73$\pm$0.03 & 2545$\pm$50 & 0.399 & 0.995\\
280 & 196$\pm$4  & 716$\pm$35 &  6.00$\pm$0.46 &  0.11$\pm$0.01 & 0.73$\pm$0.02 & 2728$\pm$52 & 0.392 & 0.996\\
290 & 188$\pm$9  & 704$\pm$71 &  6.85$\pm$1.09 &  0.10$\pm$0.01 & 0.75$\pm$0.02 & 2967$\pm$54 & 0.501 & 0.977\\
300 & 190$\pm$7  & 679$\pm$50 &  7.60$\pm$0.89 &  0.11$\pm$0.01 & 0.73$\pm$0.01 & 3133$\pm$56 & 0.495 & 0.979\\
310 & 173$\pm$3  & 649$\pm$16 &  9.33$\pm$0.42 &  0.12$\pm$0.01 & 0.68$\pm$0.01 & 3258$\pm$57 & 0.390 & 0.996\\
320 & 186$\pm$8  & 669$\pm$43 &  8.32$\pm$0.86 &  0.12$\pm$0.01 & 0.68$\pm$0.01 & 3465$\pm$59 & 0.444 & 0.990\\
330 & 178$\pm$7  & 600$\pm$34 & 10.74$\pm$1.04 &  0.07$\pm$0.01 & 0.66$\pm$0.01 & 3590$\pm$60 & 0.564 & 0.953\\
340 & 184$\pm$5  & 619$\pm$14 & 10.30$\pm$0.82 &  0.15$\pm$0.01 & 0.61$\pm$0.01 & 3554$\pm$60 & 0.398 & 0.995\\
350 & 200$\pm$6  & 658$\pm$39 &  8.73$\pm$0.89 &  0.17$\pm$0.01 & 0.58$\pm$0.01 & 3689$\pm$61 & 0.732 & 0.817\\
360 & 177$\pm$9  & 634$\pm$46 & 10.96$\pm$1.23 &  0.14$\pm$0.01 & 0.60$\pm$0.01 & 3672$\pm$61 & 0.517 & 0.972\\
\hline
\end{tabular}  
\end{table*}

\begin{figure}
\begin{center}
\includegraphics[angle=0, width=75mm, height=144mm]{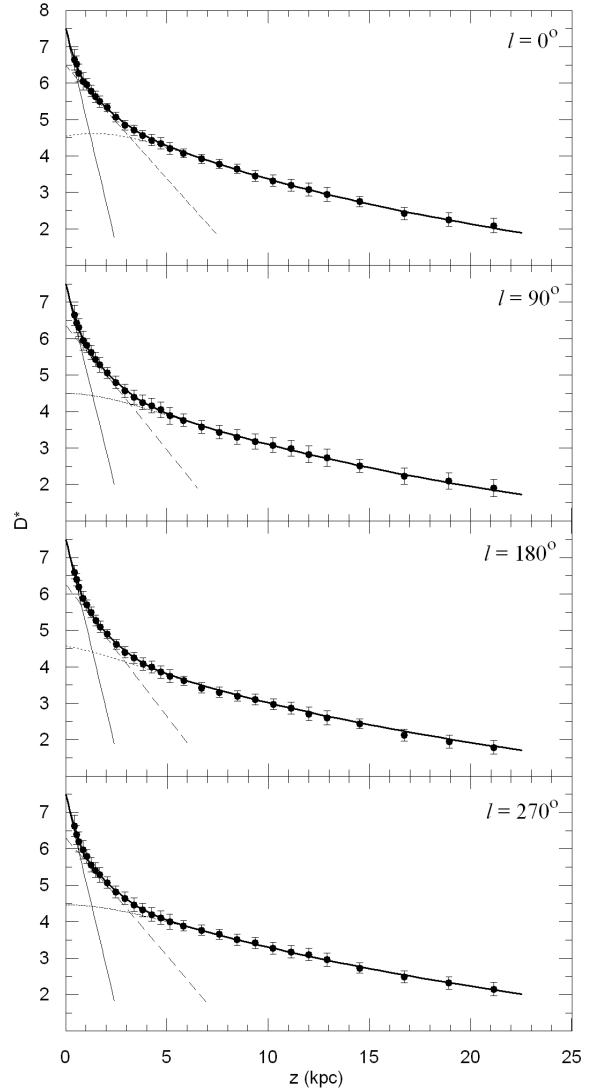}
\caption[] {Observed (symbols) and evaluated (thick solid lines) space density functions combined for stars of all three population components: thin disc (thin solid line), thick disc (dashed line) and halo (dotted line), at four different Galactic longitudes.} 
\label{SDSS_density}
\end{center}
\end {figure}

\begin{figure}
\begin{center}
\includegraphics[angle=0, width=75mm, height=108mm]{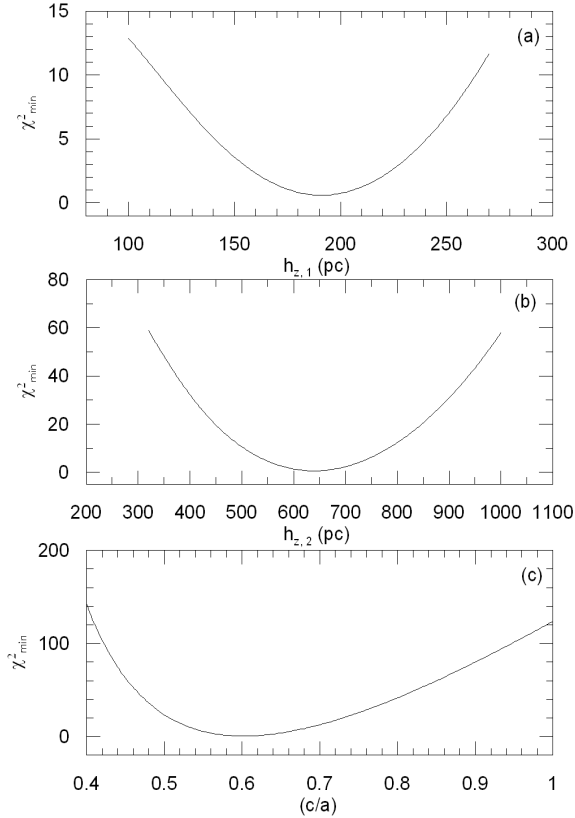}
\caption[] {The variation of $\chi^{2}_{min}$ with three Galactic model parameters. (a) with scaleheight of the thin disc, (b) with scaleheight of the thick disc, and (c) with axial ratio of the halo. The figures in the panels (a) and (b) are symmetric and the one in the panel (c) is a bit skewed to the right.} 
\label{chi-square}
\end{center}
\end {figure}

\begin{figure}
\begin{center}
\includegraphics[angle=0, width=75mm, height=128mm]{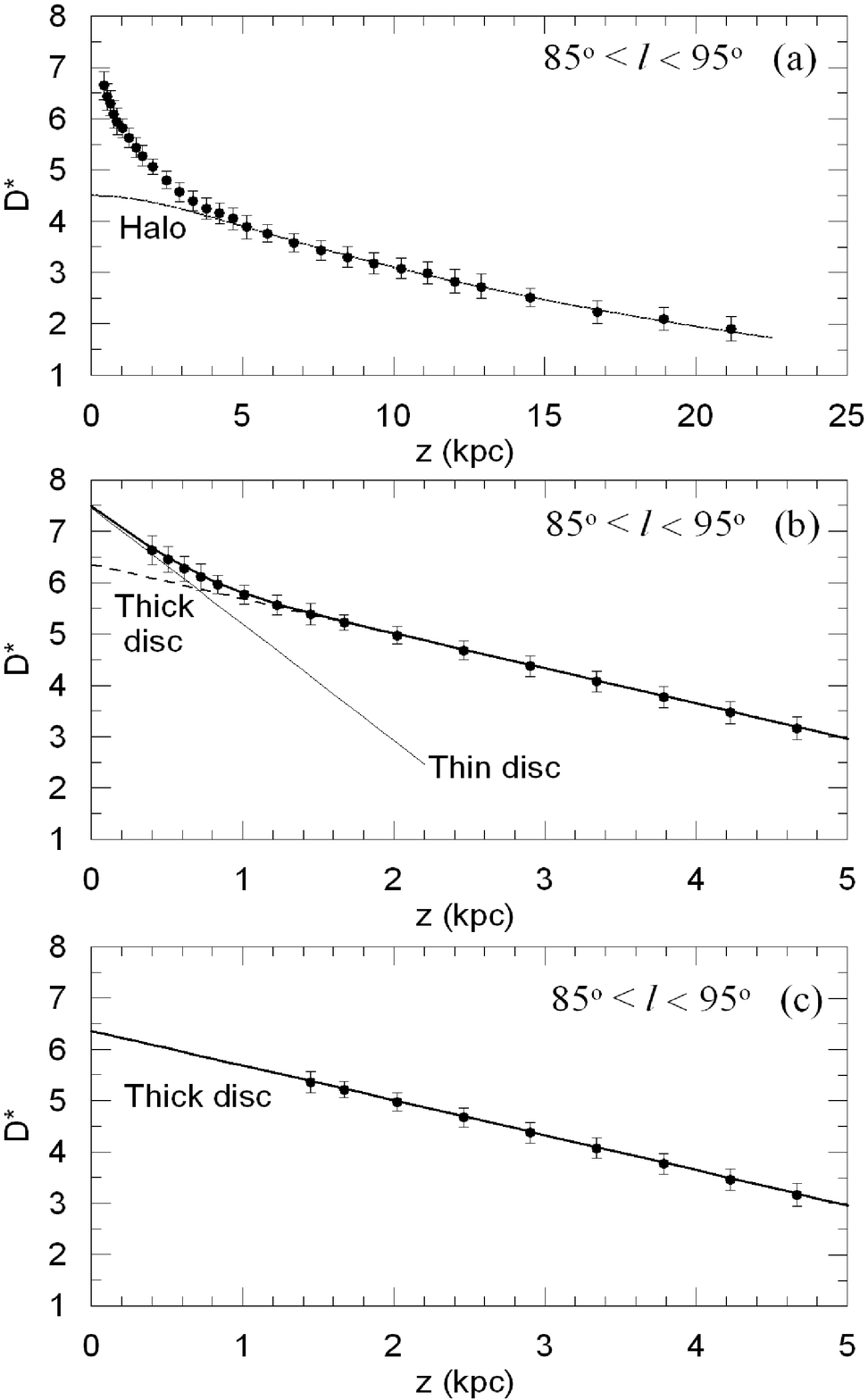}

\caption[]{An alternative procedure for the estimation of the Galactic model parameters, for the field centred at $l=90^\circ$ as an example. In panel (a), the space densities based on the observational data for distances larger than $z=5$ kpc are compared with the analytical density law of the halo. Thus, only the local space density and the axial ratio of the halo are estimated independent of the Galactic model parameters of the thin and thick discs. In panel (b), the density of the halo estimated in panel (a) is omitted, and the density function of the remaining data for $z<5$ kpc is compared with the combined analytical density laws of the thin and thick discs, which provide local space densities and scaleheights for the thin and thick discs. Finally, in panel (c) only the density function for $1.5<z\leq5$ kpc, where the thick disc is dominant, is compared with the  analytical density law of the thick disc. Thus, we estimated the model parameters of the thick disc individually. The Galactic model parameters estimated by this procedure are in agreement with the ones estimated by comparison of the combined observational based density function with the combined analytical density laws for three populations. This agreement excludes any possible degeneracy between halo and discs, and between two discs.}
\label{deg}
\end{center}
\end {figure}

We acknowledge that in this work we have not applied any correction for binarity or giant/subgiant stars. However, most of the evolved stars are probably rejected automatically due to the limiting apparent magnitude at the bright end, i.e. $g_{0}=15$. We compared the number of giants and dwarfs with apparent magnitudes $g_{0}>15$ for a reliable confirmation of our argument. We adopted a mean absolute magnitude $M_{g}=1.5$ for giants and we calculated their corresponding distance from the Galactic plane, $z=4.5$ kpc, for $g_{0}=15$. We used the local space density $D^{*}=6.35$ and the scaleheight ($H=650$ pc) for the thick disc dwarfs and we compared the corresponding space density with that of thick disc giants for local space density $D^{*}=5.58$ and the scaleheight ($H=585$ pc) adopted from \cite{Bilir06a}. It turned out that the number of thick disc giants fainter than $g_{0}=15^{m}$ is less than 8\% of the number of dwarfs. The number of halo giants relative to the number of halo dwarfs at distances $z\geq4.5$ kpc is even less, i.e. 2\%. We used the selection criteria for metal-poor giants of \cite{Helmi03} to estimate the cited number of halo giants. These authors define the location of the metal-poor giants by the following criteria: $r<19^{m}$, $1.1\leq(u-g)_{0}\leq2.0$, $0.3\leq(g-r)_{0}\leq0.8$, $-0.1<P_{1}<0.6$, $|s|>m_{s}+0.05$, where $P_{1}=0.910(u-g)_{0}+0.415(g-r)_{0}–-1.28$, $s=-0.249u+0.794g–-0.555r+0.24$ and $m_{s}=0.002$. When we apply these criteria to stars with $4<M_{g}\leq10$ we obtain 460 giants corresponding to 2\% of the total number of stars (N=19 325) for the field $85^\circ< l <95^\circ$.  

The range of the binary stars' fraction is 25--50\% depending on the spectral types of stars. If we assume a binary fraction of 50\% then the inferred scaleheight in a photometric parallax evaluation is approximately 80\% of the actual value \citep{Siegel02}. However, as we compare the scaleheight of stars of a specific population for fields in different directions of the Galaxy, we are interested only in the relative values of scaleheight but not their actual values. Hence, disregarding the binarity does not affect our results.           

\section{Galactic model parameters}     
\subsection{Estimation of the Galactic model parameters}

We estimated all the Galactic model parameters (the local space densities and scaleheights for the thin and thick discs, and the local space density and axial ratio for the halo) simultaneously, by fitting the space density functions derived from the observations (combined for the three population components) to a corresponding combination of the adopted population-specific analytical density laws. The absolutely faintest stars in this work provide space densities at short distances from the Galactic plane, $z\sim0.4$ kpc. Hence it is possible to have reliable extrapolation between them and the space density of {\em Hipparcos} in the solar neighborhood, $D^{*}=7.49$ in logarithmic form \citep{Jahreiss97}, corresponding to the mean of the local space densities for stars with $4<M_{g}\leq10$. 

We used the classical $\chi^{2}$ method for the estimation of the Galactic model parameters, a method which is made in the studies of the Galactic structure that has been used in the determination of the most recent numerical values for the Galactic model parameters \citep{Phleps00, Phleps05, Chen01, Siegel02, Du03, Du06, J08}. The comparison of the logarithmic density functions derived from the observations and the analytical density laws are given in Fig. \ref{SDSS_density}, for four fields with Galactic longitudes $l=0^\circ$, 90$^\circ$, 180$^\circ$ and 270$^\circ$. $\chi_{min}^{2}$ shows almost a symmetrical distribution, as can be observed from Fig. \ref{chi-square} which is given as an example. Hence, the errors of the Galactic model parameters could be estimated by changing a given model parameter until an increase or decrease by 1 was achieved \citep{Phleps00}. All the Galactic model parameters and their errors are given in Table \ref{tableSDSS}. The reduced $\widetilde{\chi}^{2}_{min}$ values and the corresponding probabilities are also given in Table \ref{tableSDSS}. The $\widetilde{\chi}^{2}_{min}$ values are low, whereas the probabilities are rather high confirming the reality of the Galactic model parameters. 

We also used a different procedure just to test any possible degeneracy in the estimation of the model parameters: First, we estimated the local space density and the axial ratio for the halo by comparing the logarithmic space density function for $z>5$ kpc with the analytical density law of the halo (Eq. 12). Then, we omitted the space density of the halo, estimated by the corresponding density law, and compared the new density function for $z\leq5$ kpc with the combined density laws of thin and thick discs (Fig. \ref{deg}). This procedure provides Galactic model parameters for thin and thick discs independent of the model parameters of the halo. The result of the application of the two different procedures shows that the corresponding Galactic model parameters for a specific population are either identical or they differ only by a negligible amount. Hence, we may argue that no degeneracy exists in the estimated parameters. 

A similar procedure is applied to the thin and thick discs. We estimated the solar normalizations and the scaleheights of thin and thick discs simultaneously by the space density function for $0.4<z\leq5$ kpc and compared the model parameters of the thick disc with the corresponding ones estimated by the space density function for $1.5<z\leq5$ kpc, where the thick disc is dominant. Since no significant differences could be observed between the compared parameters, we concluded that no degeneracy exists between the two discs either.  

\begin{figure*}
\begin{center}
\includegraphics[angle=0, width=160mm, height=88.88mm]{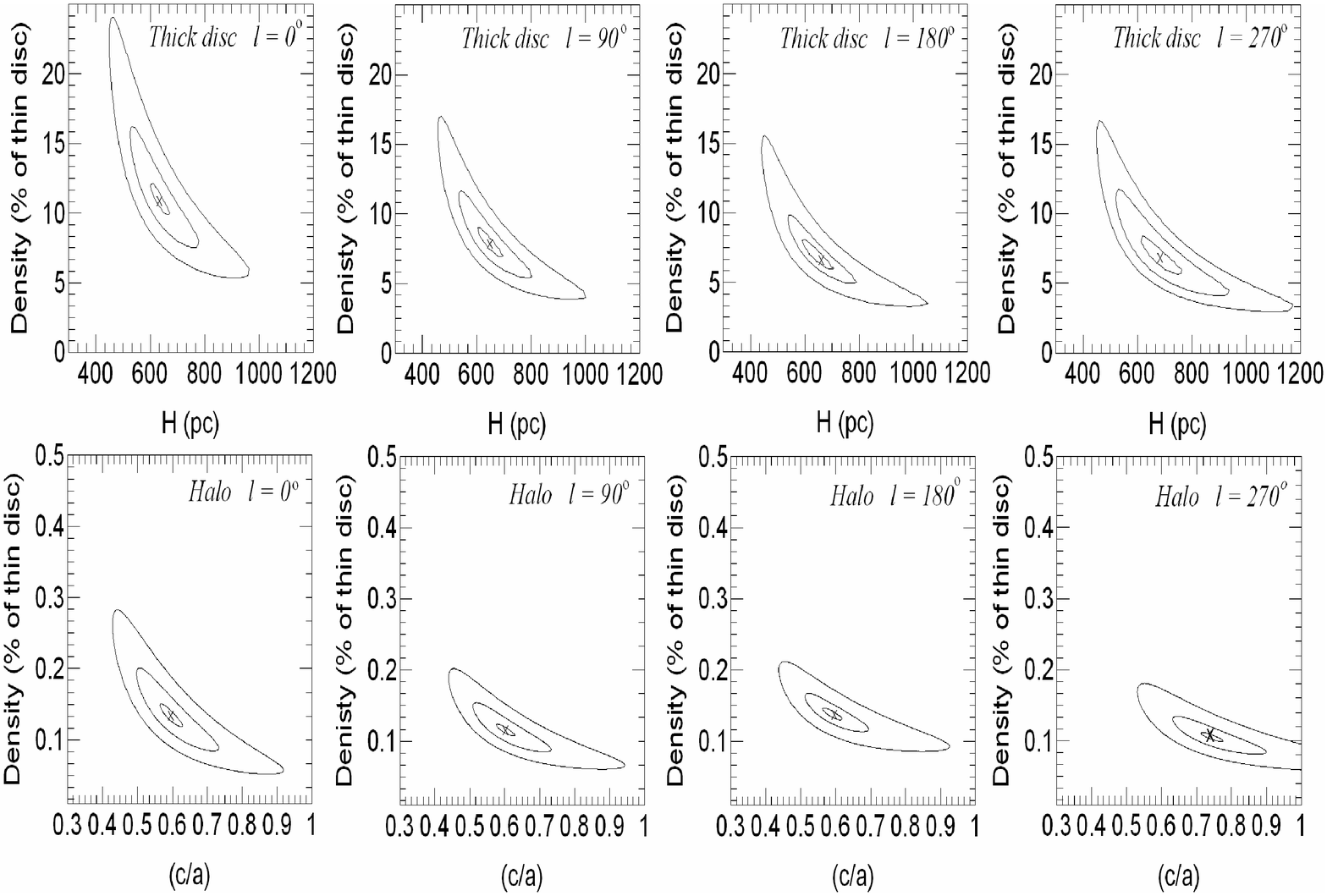}
\caption[] {Contours of equal $\chi^{2}$ obtained for different values of scaleheight and local density for thick disc, and axial ratio and local density for halo for the fields with Galactic longitudes of 0$^\circ$, 90$^\circ$, 180$^\circ$, and 270$^\circ$. The cross ($\times$) shows the minimum $\chi^{2}$ value in each panel while the contours show the 1$\sigma$, 2$\sigma$, and 3$\sigma$ confidence levels.} 
\label{chi-tk-h}
\end{center}
\end {figure*}

Additionally, we plotted the relative local space density of the thick disc versus its scaleheight, and the relative local space density of the halo versus its axial ratio, for four fields with Galactic longitudes $l=0^\circ$, 90$^\circ$, 180$^\circ$ and 270$^\circ$ in Fig. \ref{chi-tk-h} to test the same problem. The contours correspond to the same $\chi^{2}$ for $\sigma$, $2\sigma$ and $3\sigma$, where $\sigma$ is the standard deviation. In each panel, the cross shows the position for the minimum $\chi^{2}$, which defines the Galactic model parameters on the axes with accuracy.
 
\subsection{Dependence of the Galactic model parameters with the Galactic longitude}

Table \ref{tableSDSS} shows that the scaleheights of the thin and thick discs, as well as the relative local space densities of the thick disc and halo. Even the axial ratio of the halo are not the same for 36 fields. That is, these Galactic model parameters change as a function of the Galactic longitude. 

\begin{figure}
\begin{center}
\includegraphics[angle=0, width=75mm, height=34mm]{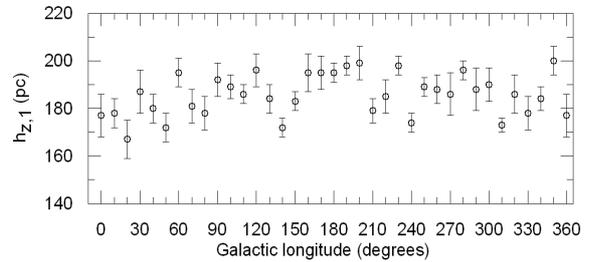}
\caption[] {Variation in the scaleheight of the thin disc with the Galactic longitude (the Galactic latitudes of the fields lie within $60^\circ\leq b \leq65^\circ$).} 
\label{TN-model}
\end{center}
\end {figure}

Fig. \ref{TN-model} shows the variation in the scaleheight of the thin disc ($h_{z}$) with the Galactic longitude ($l$). The global distribution of $h_{z}$ has a maximum at $l\approx190^\circ$, almost in the direction the anti-Galactic centre, whereas the minimum corresponds to the fields in the Galactic centre direction. However one can separate the bins into several sub-samples with segments of different slopes. Additionally, the segments corresponding to the fields with Galactic longitudes less than $150^\circ$ have negative slopes whereas the slopes of the ones with longitudes greater than $150^\circ$ are positive. 

\begin{figure}
\begin{center}
\includegraphics[angle=0, width=75mm, height=58mm]{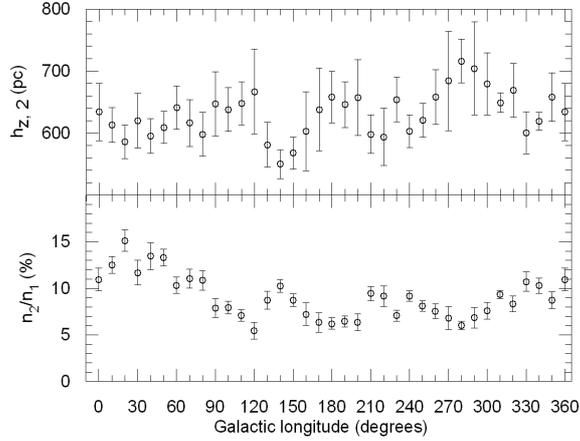}
\caption[] {Variations in the scaleheight and relative space density of the thick disc with the Galactic longitude.} 
\label{TK-model}
\end{center}
\end {figure}

\begin{figure}
\begin{center}
\includegraphics[angle=0, width=75mm, height=58mm]{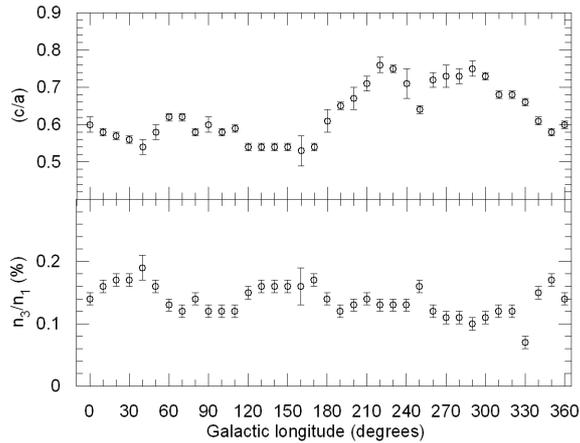}
\caption[] {Variations in the axis ratio and relative space density of the halo with the Galactic longitude.} 
\label{HH-model}
\end{center}
\end {figure}

The trend of the scaleheight of the thick disc is different than the one the thin disc (Fig. \ref{TK-model}). The maximum and minimum of the scaleheight for the global distribution are $l\approx290^\circ$ and $l\approx140^\circ$, respectively. The slopes of the segments in this figure are also different than the ones of Fig. \ref{TN-model}, for the same longitude set. The error bars are also larger than the ones for the scaleheights of the thin disc. Different trends for the relative local space densities of the thick disc ($n_{2}/n_{1}$) and halo ($n_{3}/n_{1}$), and the axial ratio of the halo ($c/a$) can be also observed in Figs. \ref{TK-model} and \ref{HH-model}. For example, the minimum of the global distribution of ($n_{2}/n_{1}$) lies within $180^\circ\leq l\leq 200^\circ$ whereas the minimum of ($n_{3}/n_{1}$) lies in an interval with larger Galactic longitudes, $270^\circ\leq l\leq 290^\circ$. We should note that one can also observe segments with different slopes in the distributions of the relative local space densities for the thick disc and halo, and of the axial ratio for the halo.

Apart from the variation of the Galactic model parameters with the Galactic longitude, one can say something about the correlation between the estimated parameters. Fig. \ref{TK-model} shows that the scaleheight of the thick disc is an increasing function of the Galactic longitude in an interval where the local space density of the thick disc is a decreasing function and vice versa. The same argument holds for the axis ratio and the local space density of the halo (Fig. \ref{HH-model}). Although there is a degeneracy between the scaleheight and the local space density of the thick disc, and between the axial ratio and the local space density of the halo (Fig. \ref{chi-tk-h}), the mentioned correlations are real, since the corresponding reduced $\widetilde{\chi}^{2}_{min}$ are rather low (Table \ref{tableSDSS}). These correlations were also cited in \cite{Buser98, Buser99} where seven fields in different directions of the Galaxy were investigated with {\em RGU} photometry.

The range, mean, and standard deviations of the Galactic model parameters are given in Table \ref{tablemodel}. The mean of the scaleheight of the thin disc is (at least) about 30$\%$ less than the one that appeared in the literature \citep{Ojha99, Buser98, Buser99, KBH04}. For the thick disc, the scaleheight and the relative local space density are close to the ones that appeared in the literature in recent years, also obtained from {\em SDSS} data. However, the upper limit of the space density is a bit higher than the cited one up until now \citep{Chen01, Siegel02}. The axial ratio and the relative local space density of the halo are rather close to the corresponding ones cited by many authors \citep {Robin96, Robin00, Chen01, Siegel02, KBH04, Bilir06c}.      

\begin{table}
\center
\caption{Equations for the variation of the different Galactic model parameters as a function of Galactic longitude (last column). 
The range, mean and the corresponding standard deviation are also given in the second, third and fourth columns, respectively.}
\label{tablemodel}
\begin{tabular}{cccc}
\hline
Parameter & Range & Mean & $s$\\
\hline
$h_{z,1}$ (pc)       & 167--200    & 186  &  9   \\
$h_{z,2}$ (pc)       & 550--716    & 631  & 37   \\
($n_{2}/n_{1}$) ($\%$) & 5.45--15.14 & 9.03 & 2.34 \\
($n_{3}/n_{1}$) ($\%$) & 0.07--0.19  & 0.14 & 0.02 \\
($c/a$)                & 0.53--0.76  & 0.63 & 0.07 \\
\hline
\end{tabular}  
\end{table}

\section{Discussion}
\subsection{Evolution of the concept about the Galactic model parameters}

Galactic researchers have been working on the modelling of our Galaxy for about 25 years. The Galactic model parameters for the discs and halo have been refined since this epoch. There is a concensus about the idea that the most refined parameters are those of the thick disc. Actually, the scaleheight of the thick disc decreased from the original value of \cite{GR83}, 1.45 kpc, to the recent one, 0.65 kpc whereas the solar normalization increased from 2 to 6--10\% \citep{Siegel02}. Despite the same density laws, different model parameters with large ranges have been cited by different researchers \citep[see Table 1 of][]{KBH04}. However, one expects Galactic model parameters either with small errors or with a short range from the recent surveys such as {\em SDSS}, {\em SEGUE}, {\em 2MASS}, {\em DENIS}, and {\em UKIDSS} which provide accurate magnitude's and colours. It seems that we must approach the problem from a physical point of view. 

We showed in \cite{KBH04} that the Galactic model parameters are absolute magnitude dependent. The errors for the model parameters estimated for a unit absolute magnitude interval are rather small, and the numerical values for a specific Galactic model parameter increases or decreases, depending on the parameter, with the absolute magnitude. For example, the range of the scaleheight of the thin disc is 264--334 pc for the absolute magnitude interval $5<M_{g}\leq13$. A second example can be given for the thick disc. The range of the relative local space density of the thick disc for the absolute magnitude interval $5<M_{g}\leq9$ is 5.25--9.77\%, coincident with the classical value given in recent works, without paying attention to the dependence of this parameter on absolute magnitude. In other words, the large range of the Galactic model parameters is unavoidable in the procedure of star counts in which separation of stars into different absolute magnitude intervals is not regarded.  

Different absolute magnitude intervals correspond to different spectral types and different populations. Stars with the brightest absolute magnitudes ($4<M_{g}\leq5$), intermediate ($5<M_{g}\leq8$), and faintest absolute magnitudes ($8<M_{g}\leq10$) are of spectral types F-G, G-K and K-M, respectively (Fig. \ref{gr-ri}). As claimed in our previous papers \citep{KBH04, Bilir06c} halo, thick disc and thin disc stars are dominant at these spectral type intervals, respectively. Hence, different Galactic model parameters estimated for different absolute magnitude diagrams correspond to different populations and hence stellar ages.  

\begin{table*}
\center
\caption{Comparison of the surface densities (number of stars per square degree) for fields in different quadrants as a function of colour and apparent magnitude. The suffix denotes the longitude of the field in question.}
\label{ratio}
\begin{tabular}{ccccccc}
\hline
$(g-r)_{0}$ & $g_{1}-g_{2}$  &$N_{30}/N_{330}$ & $N_{60}/N_{300}$ & $N_{90}/N_{270}$ & $N_{120}/N_{240}$ & $N_{150}/N_{210}$ \\
\hline
$(g-r)_{0}\leq0.35$     & $15<g_{0}\leq16$ & 1.55 &       0.88 &       0.91 &       0.83 &       1.18 \\
$(g-r)_{0}\leq0.35$     & $16<g_{0}\leq17$ & 0.88 &       1.01 &       0.83 &       0.95 &       1.00 \\
$(g-r)_{0}\leq0.35$     & $17<g_{0}\leq18$ & 0.90 &       0.84 &       0.88 &       0.82 &       0.91 \\
$(g-r)_{0}\leq0.35$     & $18<g_{0}\leq19$ & 0.87 &       0.88 &       0.80 &       0.75 &       0.85 \\
$(g-r)_{0}\leq0.35$     & $19<g_{0}\leq20$ & 0.80 &       0.72 &       0.67 &       0.74 &       0.75 \\
$(g-r)_{0}\leq0.35$     & $20<g_{0}\leq21$ & 0.69 &       0.58 &       0.58 &       0.53 &       0.57 \\
$(g-r)_{0}\leq0.35$     & $21<g_{0}\leq22$ & 0.47 &       0.57 &       0.56 &       0.38 &       0.45 \\
\hline
$0.35<(g-r)_{0}\leq0.60$& $15<g_{0}\leq16$ & 1.16 &       1.08 &       0.99 &       0.95 &       1.02 \\
$0.35<(g-r)_{0}\leq0.60$& $16<g_{0}\leq17$ & 1.14 &       1.15 &       1.04 &       0.96 &       0.92 \\
$0.35<(g-r)_{0}\leq0.60$& $17<g_{0}\leq18$ & 1.11 &       1.16 &       0.94 &       0.84 &       0.92 \\
$0.35<(g-r)_{0}\leq0.60$& $18<g_{0}\leq19$ & 1.01 &       0.94 &       0.83 &       0.78 &       0.78 \\
$0.35<(g-r)_{0}\leq0.60$& $19<g_{0}\leq20$ & 0.83 &       0.82 &       0.74 &       0.71 &       0.75 \\
$0.35<(g-r)_{0}\leq0.60$& $20<g_{0}\leq21$ & 0.75 &       0.71 &       0.69 &       0.60 &       0.74 \\
$0.35<(g-r)_{0}\leq0.60$& $21<g_{0}\leq22$ & 0.61 &       0.67 &       0.67 &       0.57 &       0.67 \\
\hline
$0.60<(g-r)_{0}\leq1.20$& $15<g_{0}\leq16$ & 1.04 &       1.06 &       1.02 &       0.97 &       0.93 \\
$0.60<(g-r)_{0}\leq1.20$& $16<g_{0}\leq17$ & 1.26 &       1.16 &       0.91 &       0.88 &       0.95 \\
$0.60<(g-r)_{0}\leq1.20$& $17<g_{0}\leq18$ & 1.15 &       1.18 &       0.98 &       0.95 &       0.91 \\
$0.60<(g-r)_{0}\leq1.20$& $18<g_{0}\leq19$ & 1.06 &       1.11 &       1.03 &       0.88 &       0.89 \\
$0.60<(g-r)_{0}\leq1.20$& $19<g_{0}\leq20$ & 1.03 &       1.07 &       0.97 &       0.88 &       0.90 \\
$0.60<(g-r)_{0}\leq1.20$& $20<g_{0}\leq21$ & 0.98 &       1.02 &       0.92 &       0.80 &       0.89 \\
$0.60<(g-r)_{0}\leq1.20$& $21<g_{0}\leq22$ & 0.93 &       0.88 &       0.83 &       0.79 &       0.91 \\
\hline
$(g-r)_{0}>1.20$        & $15<g_{0}\leq16$ & 0.90 &       1.17 &       0.83 &       0.76 &       1.05 \\
$(g-r)_{0}>1.20$        & $16<g_{0}\leq17$ & 1.14 &       1.12 &       0.98 &       1.07 &       0.95 \\
$(g-r)_{0}>1.20$        & $17<g_{0}\leq18$ & 1.17 &       1.22 &       1.18 &       0.90 &       0.90 \\
$(g-r)_{0}>1.20$        & $18<g_{0}\leq19$ & 1.16 &       1.20 &       1.13 &       1.01 &       0.92 \\
$(g-r)_{0}>1.20$        & $19<g_{0}\leq20$ & 1.19 &       1.23 &       1.06 &       0.98 &       0.94 \\
$(g-r)_{0}>1.20$        & $20<g_{0}\leq21$ & 1.20 &       1.26 &       1.15 &       0.99 &       0.89 \\
$(g-r)_{0}>1.20$        & $21<g_{0}\leq22$ & 1.14 &       1.28 &       1.10 &       0.98 &       0.86 \\
\hline
\end{tabular}  
\end{table*}

\subsection{Interpretation of the dependence of the Galactic model parameters with the Galactic longitude}
\subsubsection{Scenarios for the asymmetric structure of the Galaxy and their confirmation by star counts}

The differences between the numerical values of a given Galactic model parameter estimated in different directions of the Galaxy (mainly at large galactocentric distances), could be explained by the overdensity regions with respect to an axisymmetric halo. Two competing scenarios have been proposed for this: the first one is concerned with the triaxiality of the halo \citep{Newberg06, Xu06, J08}, whereas the second one is related to the remnants of some historical events \citep[cf.][]{Wyse05}. Although \citet{Newberg05} claimed that the thick disc was symmetric about the Galactic longitude $l=180^\circ$; \citet{Parker03} interpretered the excess in the numbers of blue- and -intermediate- coloured stars above and below the Galactic plane in quadrant I as the asymmetric structure of the thick disc. \citet{Parker03} propose similar scenarios for their observations: a) fossil remnant of a merger, b) a triaxial thick disc or halo, and c) interaction of the thick disc/inner halo stars with the bar in the disc. The limiting magnitude in our work is much fainter ($g_{0}=22$) than the limiting magnitude (O=18) of the authors just mentioned. Hence, we thought we can obtain more reliable results, if we use a similar procedure. 

\begin{figure}
\begin{center}
\includegraphics[angle=0, width=75mm, height=41.25mm]{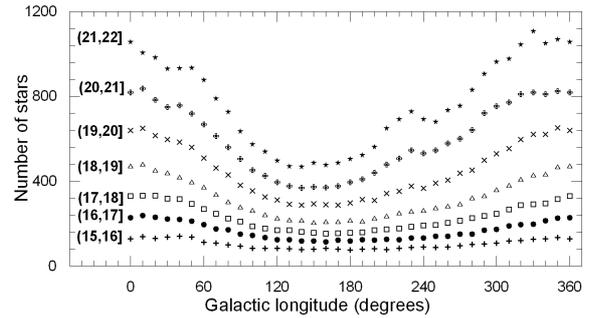}
\caption[] {Variation of the number of stars with different $g_{0}$ apparent magnitudes as a function of the Galactic longitude.} 
\label{starcount-magnitude}
\end{center}
\end{figure}

We adopted the procedure of \citep{Newberg05}, and we plotted the surface density (number of stars per square degree of a field) as a function of Galactic longitude, for different apparent $g$- magnitude intervals, i.e. (15, 16], (16, 17], (17, 18], (18, 19], (19, 20], (20, 21], and (21, 22] (Fig. \ref{starcount-magnitude}). An asymmetric structure can be observed not only for the faint magnitudes which correspond to the halo stars, but also for the intermediate magnitudes that favour the thick disc stars. All the functions have a flat minimum in the interval $120^\circ<l<180^\circ$, which is more conspicuous at fainter magnitudes. That is, there is a deficiency in the number of stars in that longitude interval. For example, the surface density in this interval is about 50\% less than the one for a field in the Galactic centre direction, for the apparent magnitude interval $21<g_{0}<22$. The less surface density for the fields in the anti-centre direction relative to the fields in the Galactic centre direction is an effect of the disc scalelength. However, this effect should be most efficient at the field with longitude $l=180^\circ$ which is not the case in Fig. \ref{starcount-magnitude}. The excess of stars for the fields $l\approx230^\circ$ corresponds to the ``tail of the Sagittarius tidal stream'' overdensity region cited by \citet{Newbergetal06}, that was also noted by \citet{Xu06}.

Following \citet{Parker03}, we compared the number of stars for five pairs of fields which are symmetric relative to the meridian, i.e. the plane perpendicular to the Galactic plane and passing through the centre of the Galaxy and the Sun. The sample pairs are: (30$^\circ$, 330$^\circ$), (60$^\circ$, 300$^\circ$), (90$^\circ$, 270$^\circ$), (120$^\circ$, 240$^\circ$), and (150$^\circ$, 210$^\circ$). The comparison is carried out for four colours, i.e. $(g-r)_{0}\leq0.35$, $0.35<(g-r)_{0}\leq0.60$, $0.60<(g-r)_{0}\leq1.20$, and $1.20<(g-r)_{0}$. As mentioned in the Introduction, blue stars in the range $15<g_{0}<18$ are dominated by thick-disc stars with a turn-off $(g-r)_{0}\approx0.33$, while halo stars become significant for $g_{0}>18$, with a turn-off $(g-r)_{0}\approx0.2$. Red stars, $(g-r)_{0}\approx1.3$ are dominated by thin disc stars at all apparent magnitudes. We shall treat the ratio of stars given in Table \ref{ratio} according to their colours. We will use the notation $N_{i}/N_{j}$ for the ratio of stars, for simplicity, where suffixes $i$ and $j$ denote the Galactic longitudes of the fields in question.
\par
(a) The colour range $(g-r)_{0}\leq 0.35$\\ 
Apparently faint stars in this interval are halo stars, whereas the bright ones are (thin or thick) disc stars. The ratio of the number of stars in quadrant I to the number of stars in quadrant IV is about 1 or greater than 1 for bright apparent magnitudes, whereas it decreases when one goes to the fainter magnitudes, and it approaches about 0.5 at $21<g_{0}\leq22$. The inequality of number of stars for the halo population is due to the triaxiality of this component of the Galaxy. The same case holds for the halo stars in quadrant II and quadrant III. However, for bright stars, the interpretation of the ratio of stars seems difficult (but see the following sections).
\par
(b) The colour range $0.35<(g-r)_{0}\leq 0.60$\\ 
This is a perfect colour range for the interpretation of the ratios of the number of stars in the corresponding fields, since relatively bright stars belong to the thick disc population whereas the faint ones are halo population objects. The ratios $N_{30}/N_{330}$ and $N_{60}/N_{300}$ are greater than 1 for three apparently bright intervals, i.e. $g_{0}$: (15, 16], (16, 17] and (17, 18] whereas it is less than 1 for $g_{0}$: (19, 20], (20, 21] and (21, 22]. An excess in number of stars in quadrant I relative to quadrant IV can be explained by the existence of a bar with its near end at the Galactic longitude $l\approx27^\circ$ \citep{Parker03}. According to this argument one expects a deficiency of stars in quadrant II relative to quadrant III. Actually, this is the case in our work, i.e. $N_{120}/N_{240}$ and $N_{150}/N_{210}$ are less than 1 for (thick) disc stars. Here also, the inequality of number of stars with apparently faint magnitudes is due to the triaxiality of the halo. The apparent magnitude interval $18<g_{0}\leq19$ is a transition region between disc and halo populations. 
\par
(c) The colour range $0.60<(g-r)_{0}\leq 1.20$\\ 
This is also a perfect colour range for discussing the number of stars in the fields in different quadrants. The ratios $N_{30}/N_{330}$ and $N_{60}/N_{300}$  are greater than 1 for five apparent magnitude intervals i.e. $g_{0}$: (15, 16], (16, 17], (17, 18], (18, 19] and (19, 20], indicating a fainter limiting apparent magnitude for the thick-disc stars in this colour range. As expected, the ratios $N_{120}/N_{240}$ and $N_{150}/N_{210}$ are less than 1 for the same apparent magnitude intervals. The combination of these results confirm our argument stated in the previous section, i.e. the inequality of the number of stars brighter than $g_{0}=20$ in quadrant I and quadrant IV originates from the effect of the disc bar. For stars with $g_{0}>20$, all the ratios of the number of stars mentioned above are less than 1. These stars belong to the population of a triaxial halo, as cited above. We should note that the ratio of the number of faint stars (halo stars) with this colour range is larger than the corresponding ones in the previous colours.
\par
(d) The colour range $(g-r)_{0}>1.20$\\
As cited in the previous sections, the red stars belong to the thin disc population. Since the number of stars is rather small, we omit the brightest apparent magnitude interval, (15, 16]. For the other six intervals, the ratios $N_{30}/N_{330}$ and $N_{60}/N_{300}$ are greater than 1, as in the previous sections. On the other hand, the ratios $N_{120}/N_{240}$ and $N_{150}/N_{210}$ are less than 1 or equal to 1. That is, although the asymmetrical structure of the disc can also be explained  by the data in this section, there is a slight difference between the effects of the bar on the stars in quadrant I and quadrant III. 

\subsubsection{Relation between the variation of the Galactic model parameters and the asymmetric structure of the Galaxy}
The global distributions of the Galactic model parameters in Figs. \ref{TN-model} - \ref{HH-model} show minimum and maximum values at different Galactic longitudes, whereas the investigation of these distributions with a short scale reveal interesting sub-structures, i.e. one can separate the plots in a figure into several sub-sets which form segments with different slopes. It is difficult to interpretate the dependence of the Galactic model parameters by means of such a distribution. However, this can be done by combining the observed distributions in Figs. \ref{TN-model} - \ref{HH-model} and the results in Section 4.2.1. The difference between the corresponding Galactic model parameters of the halo for different Galactic longitudes confirm the argument that the halo has a triaxial structure. However, for the discs the reason is different. Different surface densities between the fields in quadrants I and IV; and in quadrants III and II confirm the effect of a disc bar. The bar induces a gravitational ``wake'', traps and piles up stars behind it \citep{Hernquist92, Debattista98}. Thus, in response to a bar, one expects an excess of stars in quadrant I (quadrant III) over quadrant IV (quadrant II) which is the case in our work. Also, the disc flarings produce a change in the scaleheight of the thin and thick discs as the galactocentric distance varies. This effect is well explained in \citet{L02, L04}, and \citet{Cabrera07}, who showed that the thick disc presents a flaring in the opposite sense as the one for the thin disc, that is, a decrease in $h_{z}$ as $R$ increases for the thick disc, whereas there is an increase in $h_{z}$ with $R$ for the thin disc \citep{L02, Momany06}. 
 
\begin{figure}
\begin{center}
\includegraphics[angle=0, width=80mm, height=94.1mm]{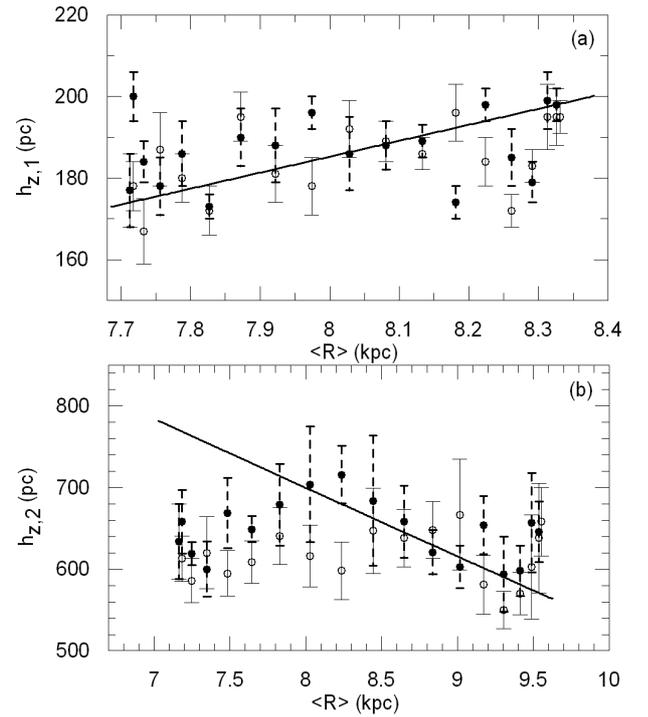}
\caption[] {Variation of the scaleheight of the thin (a) and thick (b) discs with 
$<R>$. The symbols ($\circ$) and ($\bullet$) correspond to the data for the fields with 
Galactic longitudes $0^{\circ}<l\leq180^{\circ}$ and $180^{\circ}<l\leq360^{\circ}$, 
respectively. The error bars for the fields with $180^{\circ}<l\leq360^{\circ}$ are 
marked with a different symbol (dashed segment) in order to separate them from the 
error bars of the fields with $0^{\circ}<l\leq180^{\circ}$.} 
\label{RvsTNTK}
\end{center}
\end {figure}

Fig. \ref{RvsTNTK} shows the trend in both the thin and thick disc scaleheights with the galactocentric distance. In the upper panel, the solid line represents the change in the scaleheight of the thin disc obtained by \citet{L02}, \citep[eq. 4 in][]{Cabrera07}, with $h_{z}$ ($R_{\odot}$) = 185 pc. It is clear that the results obtained are compatible with this flaring, but the scatter is too high to be conclusive in this statement. Also, the range of galactocentric distances covered by the data is too small to infer a proper fit from the data. In the lower panel, the solid line shows the relationship found in \citet[][their eq. 5]{Cabrera07} by using {\em 2MASS} data at latitudes $<b>=65^\circ$, but now considering $h_{z}$ ($R_{\odot}$) = 700 pc \footnote{both scaleheights in the solar neighbourhood are dependent on the populations under consideration, so a slight decrease in those values with respect to that obtained for the giant population, dominant in the {\em 2MASS} data, is not unexpected}. Data for $R>8$ kpc are  compatible with the trend previously obtained, although there is a decrease in the values for the innermost data with respect to the predictions of this law that would need a further explanation. Therefore, it seems that the observed trend in the scaleheight of the thick disc via the {\em 2MASS} data is also reproducible with the {\em SDSS} data. Hence the disc flarings have an important contribution to the observed changes in the parameters with galactic longitude, as different longitudes imply different ranges of galactocentric distances.

The two points for each $<R>$ bin shown in Fig. \ref{RvsTNTK} correspond to the different scaleheights of the two symmetric fields relative to the meridian. This confirms the asymmetric structure of the thin (upper panel) and thick (lower panel) discs. The different behaviours of the radial variation of the scaleheights of thin and thick discs allow specific predictions about the radial dependence of the vertical velocity dispersion of the discs, since at a given $<R>$ they are in the same gravitational field. The different behaviours are real, and, as cited at the end of the first paragraph of this section, they are due to the opposite sense of the flaring of two discs, i.e. $h_{z}$ decreases as $R$ increases for the thick disc, whereas there is an increase in $h_{z}$ with $R$ for the thin disc. The variations of the scaleheights of thin and thick discs with Galactic longitude are also different (Fig. \ref{TN-model}  and the upper panel of Fig. \ref{TK-model}). Additionally, one can see more easily that the scaleheights of thin or thick discs for two fields that are symmetric relative to the meridian are different which, again, confirms the asymmetrical structure of the two discs.
\par 
Conclusion: The dependence of the Galactic model parameters on the Galactic longitude can be explained as a result of the combined effects of the triaxial structure of the halo, the gravitational effect of the disc bar, and the thin and thick disc flarings, which are more dominant in some Galactic coordinates with respect to others.

\section{Acknowledgment}
Thanks are given to an anonymous referee, who made very important suggestions 
that have improved the overall quality of the work presented in this paper. 
This work was supported by the Research Fund of the University of Istanbul: 
Project number: BYP 941-02032006. We thank Hikmet \c{C}akmak and Tu\u{g}kent 
Akkurum for preparing some computer codes for this study. S. Karaali thanks 
to T.C. Beykent University for financial support.  

The {\em SDSS} is managed by the Astrophysical Research Consortium (ARC) for 
the Participating Institutions. The Participating Institutions are The 
University of Chicago, Fermilab, the Institute for Advanced Study, the Japan 
Participation Group, The Johns Hopkins University, the Korean Scientist Group, 
Los Alamos National Laboratory, the Max-Planck-Institute for Astronomy (MPIA), 
the Max-Planck-Institute for Astrophysics (MPA), New Mexico State University, 
University of Pittsburgh, University of Portsmouth, Princeton University, the 
United States Naval Observatory, and the University of Washington. Funding for 
the creation and distribution of the {\em SDSS} Archive has been provided by 
the Alfred P. Sloan Foundation, the Participating Institutions, the National 
Aeronautics and Space Administration, the National Science Foundation, the 
U.S. Department of Energy, the Japanese Monbukagakusho, and the Max Planck 
Society. The {\em SDSS} Web site is http://www.sdss.org/.

\begin{appendix}

\section{Two colour diagram}
\begin{figure}
\begin{center}
\includegraphics[scale=0.35, angle=0]{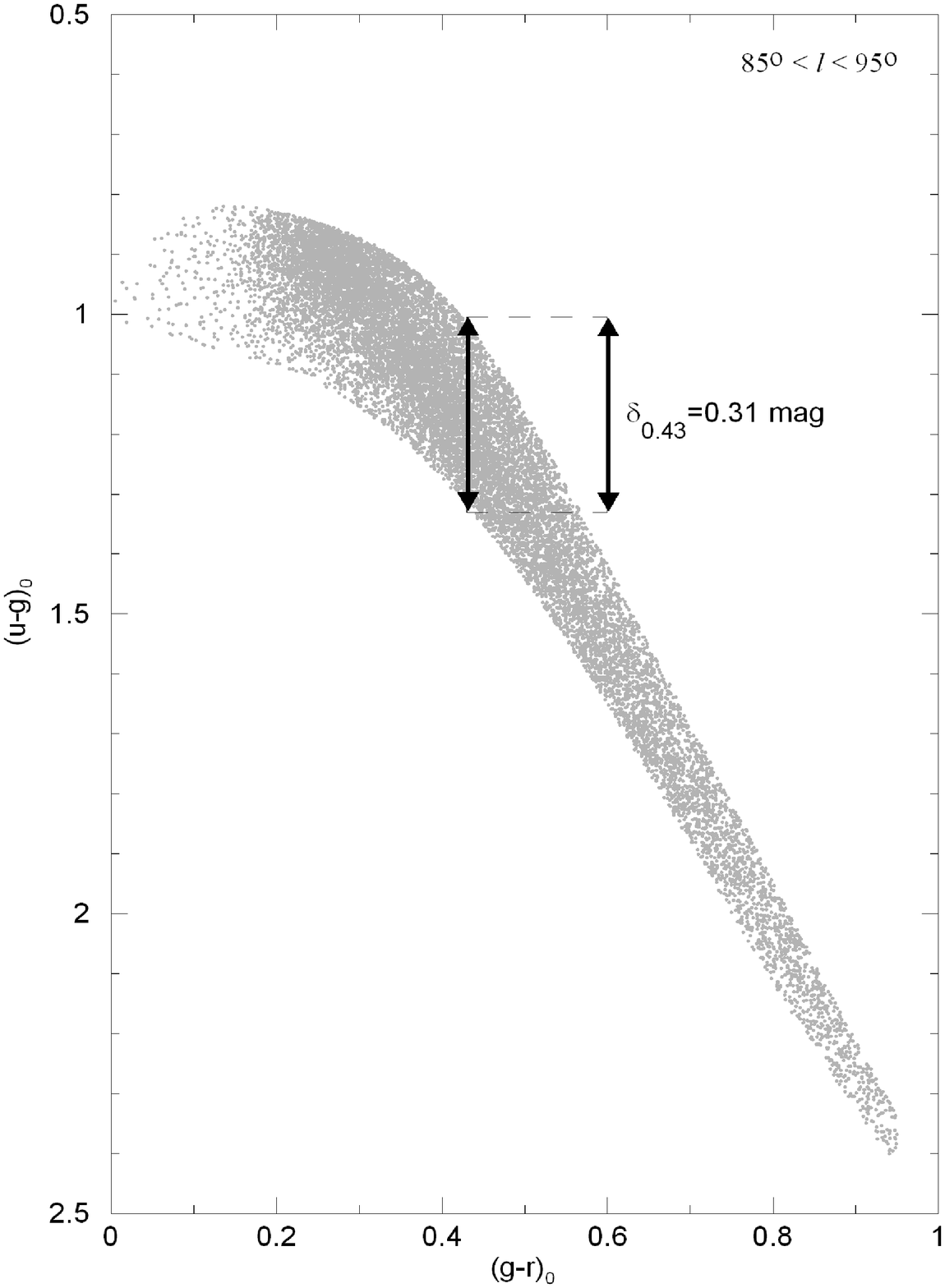}
\caption{$((u-g)_{0}, (g-r)_{0})$ two--colour diagram for stars in the field 
with longitude $l=90^{0}$. The difference in UV-–excess between the upper and lower 
envelopes of the diagram is small at the red end of the diagram which causes an 
uncertainty in the metallicity estimation. The maximum UV--excess, $\Delta 
(u-g)=0.31$ mag, corresponds to the colour $(g-r)_{0}=0.43$ mag which is equivalent 
to $(B-V)_{0}=0.60$ mag in the {\em UBV} photometry.} 
\label{two-colour}
\end{center}
\end{figure}

\end{appendix}

\end{document}